\documentclass[12pt]{article}

\usepackage{xspace}
\usepackage{epsfig}
\usepackage{epsf}
\usepackage{latexsym}
\usepackage{amsmath}
\usepackage{times}

\usepackage{amssymb}
\usepackage{epsfig}
\usepackage{psfrag}

\newcommand{\ignore}[1]{ }

\newcommand{\MST}{\mathit{MST}}
\newcommand{\OPT}{\mathit{OPT}}

\newcommand{\opt}{\mathit{opt}}

\def\qed{ \ \vrule width.2cm height.2cm depth0cm\smallskip}

\def\proof{\noindent {\bf Proof. \ }}

\newtheorem{proposition}{Proposition}
\newtheorem{lemma}[proposition]{Lemma}
\newtheorem{fact}[proposition]{Fact}
\newtheorem{theorem}[proposition]{Theorem}

\def\hQ{{\hat{Q}}}

\def\tT{{\tilde{T}}}

\newcommand{\cB}{{\cal B}}

\newcommand{\cA}{{\cal A}}
\newcommand{\ZZ}{{\mathbb{Z}}}

\newcommand{\eps}{\epsilon}

\begin{document}

\title{1.85 Approximation for Min-Power Strong Connectivity}
\author{ {G. Calinescu}
\thanks{Department of Computer Science,
Illinois Institute of Technology, Chicago, IL 60616, USA. 
{\tt calinescu@iit.edu}. Research supported in part by NSF grants
CCF-0515088 and NeTS-0916743, and the ERC StG project PAAl no. 259515.
}
}
\date{}
\maketitle

\begin{abstract}
Given a directed
simple graph $G=(V,E)$ and a cost function $c:E \rightarrow R_+$,
the {\em power} of a vertex $u$ in a directed spanning subgraph $H$ is given by
$p_H(u) = \max_{uv \in E(H)} c(uv)$, and corresponds to the
energy consumption required for wireless node $u$ to transmit to all nodes
$v$ with $uv \in E(H)$.
The {\em power} of $H$ is given by $p(H) = \sum_{u \in V} p_H(u)$.

Power Assignment seeks to minimize $p(H)$ while $H$
satisfies some connectivity constraint. In this paper, we assume
$E$ is bidirected (for every directed edge $e \in E$, the opposite edge exists
and has the same cost), while $H$ is required to be strongly connected.
This is the original power assignment problem introduced by Chen and Huang
in 1989,
who proved that a bidirected minimum spanning tree
has approximation ratio at most 2 (this is tight).
In Approx 2010, we introduced a Greedy approximation algorithm 
and claimed a ratio of $1.992$.  Here we improve the analysis to $1.85$,
combining techniques from Robins-Zelikovsky (2000) for Steiner Tree,
and Caragiannis, Flammini, and Moscardelli (2007) for the broadcast version
of Power Assignment, together with a simple idea inspired by
Byrka, Grandoni, Rothvo\ss, and Sanit\`{a} (2010).

The proof also shows that a natural linear programming relaxation,
introduced by us in 2012, has the same $1.85$ integrality gap.
\end{abstract}

\section{Introduction}

There has been a surge of research in Power Assignment problems since 2000
(among the earlier papers are
\cite{ramanathan00topology,WLBW01,HajiaghayiIM03})
This class of problems take as input a directed
simple graph $G=(V,E)$ and a cost function $c:E \rightarrow R_+$.
The {\em power} of a vertex $u$ in a directed spanning simple 
subgraph $H$ of $G$ is given by
$p_H(u) = \max_{uv \in E(H)} c(uv)$, and corresponds to the
energy consumption required for wireless node $u$ to transmit to all nodes
$v$ with $uv \in E(H)$.
The {\em power} (or {\em total power})
of $H$ is given by $p(H) = \sum_{u \in V} p_H(u)$.

The study of the min-power power assignment 
was started by Chen and Huang \cite{CH89}, which consider, as we do,
the case when $E$ is bidirected,
(that is, $uv \in E$ if and only if $vu \in E$, and if weighted, the two
edge have the same cost;
this case was sometimes called ``symmetric" or  ``undirected" in the literature)
while $H$ is required to be strongly connected. 
We call this problem {\em Min-Power Strong Connectivity}.
We use with the same name both the (bi)directed and 
the undirected version of $G$.
\cite{CH89} prove that the bidirected version
of a minimum (cost) spanning tree (MST)
of the input graph $G$ has power at most twice the optimum,
and therefore the MST algorithm has approximation ratio at most 2.
This is known to be tight (see Section \ref{s_prel}).

We improve this to $1.85$ 
by combining techniques from Robins-Zelikovsky 
\cite{RZ00,RobinsZ05} for Steiner Tree, 
Caragiannis, Flammini, and Moscardelli \cite{CFM07}
for the broadcast version of symmetric Power Assignment
(assuming a bidirected $G=(V,E,c)$ and a ``root" $u \in V$ is given,
$H$ must contain a directed path from $u$ to every vertex of $G$),
together with a simple idea inspired by
Byrka, Grandoni, Rothvo\ss, and Sanit\`{a} (2010).

Very restricted versions of Min-Power Strong Connectivity
have been proven NP-Hard \cite{KKKP00,CPS00,CK07}. Other than \cite{C10},
we are not aware
of better than a factor of 2 approximation except for \cite{CK07},
(where $c:E \rightarrow \{A,B\}$, for $0 \leq A < B$; see also \cite{C10}),
\cite{CCPRS05} (where $c$ is assumed to be a metric), and
the exact (dynamic programming) algorithms of
\cite{KKKP00} for the specific case
where each vertex of $G$ maps to a point on a line, and $c(uv)$ is an
increasing function of the Euclidean distance 
between the images of $u$ and $v$.
A related version, also NP-Hard, asks for $H$ to be bidirected 
(also called ``undirected" or ``symmetric" in previous papers).
This problem is called Min-Power Symmetric Connectivity, 
and the best known ratio of $5/3 + \eps$ \cite{ACMPTZ06} is obtained with
techniques first applied to Steiner Tree;
when $c:E \rightarrow \{A,B\}$ one gets $3/2$ with the same method
\cite{NY09}. In fact, many but not all power assignment 
algorithms use techniques from Steiner Tree variants
(or direct reduction to Steiner Tree variants; these connections 
to Steiner Tree are not obvious and cannot be easily explained), 
and in particular
Caragiannis et al \cite{CFM07} uses the relative greedy
heuristic of Zelikovsky \cite{Z96}. New interesting techniques were also
developed for power assignment problems, as in
\cite{KortsarzMNT08}, an improvement over \cite{HajiaghayiKMN07}.

The existing lower bound of the optimum, which we use,
is the cost of the minimum spanning tree of $G$. Indeed 
(argument from \cite{CH89}), 
the optimum solution $\OPT$ contains an 
in-arborescence rooted at $v$, for some $v \in V$, and then, 
for all $u \in  V \setminus \{v\}$, $p_{\OPT}(u)$ is at least
the cost of the directed edge connecting 
$u$ to its parent in this in-arborescence, whose total cost 
is at least the cost of the minimum spanning tree of $G$.

We also use a relative greedy method as in \cite{Z96,RZ00};
Robins-Zelikovsky \cite{RZ00} is rarely used as a technique, 
and not by only citing the ratio (improved by now in \cite{BGRS10}).
We use the natural structures  of \cite{CFM07} to improve over the
minimum spanning tree: these are {\em stars}, directed trees of height 1. 
Our second lower bound (improved over \cite{C10})
comes from ``covering"
the edges of the a spanning tree by the stars of the optimum solution.
With precise definitions later, we just mention that
an edge of a tree is covered by a star if it is on a path of the tree between
two vertices of the star.
A ``cheap" fractional covering can be easily obtained from either optimum
or the linear programming relaxation.
In our earlier work \cite{C10}, we used an integer cover which was
extremely hard to obtain. 
Using fractional covers (inspired by \cite{BGRS10})
is the only significant difference of this version versus \cite{C10}.
Also, interestingly, the 
submodularity of the covering function is only used implictly.

\section{Preliminaries}
\label{s_prel}

In directed graphs, we use {\em arc} to denote a directed edge.
In a directed graph $K$,
an {\em incoming arborescence}  rooted at $x \in V(K)$ is a 
spanning subgraph $T$ of $K$
such that the underlying undirected graph of $T$ is a tree 
and every vertex of $T$ other than $x$ has exactly one outgoing arc in $T$. 

Given an arc  $xy$, its {\em undirected version} is 
the undirected edge with endpoints $x$ and $y$.
Arcs $xy$ and $yx$ are {\em antiparallel}, and the antiparallel arcs
resulting from undirected edge $uv$ are $uv$ and $vu$; if undirected edge
 $uv$ has cost, then each of the two antiparallel arcs
resulting from undirected edge $uv$  have this cost.
We sometimes identify a spanning tree $T$ with its set of edges.

An alternative definition of our problem (how it was originally posed) is:
we are given a simple undirected graph $G=(V,E)$
and a cost function $c:E \rightarrow R_+$.
A power assignment is a function $p:V \rightarrow R_+$, and it
induces a simple directed graph $H(p)$ on vertex set $V$ given by
$xy$ being an arc of $H(p)$ if and only if
 $\{x,y\} \in E$ and $p(x) \geq c(\{x,y\})$.
The problem is to minimize $\sum_{u \in V} p(u)$ subject to $H(p)$
being strongly connected.
To see the equivalence of the definition, given directed spanning subgraph 
$H$, define for each $u \in V$ the power assignment $p(u) = p_H(u)$.

\begin{figure}
\centerline{\psfig{figure=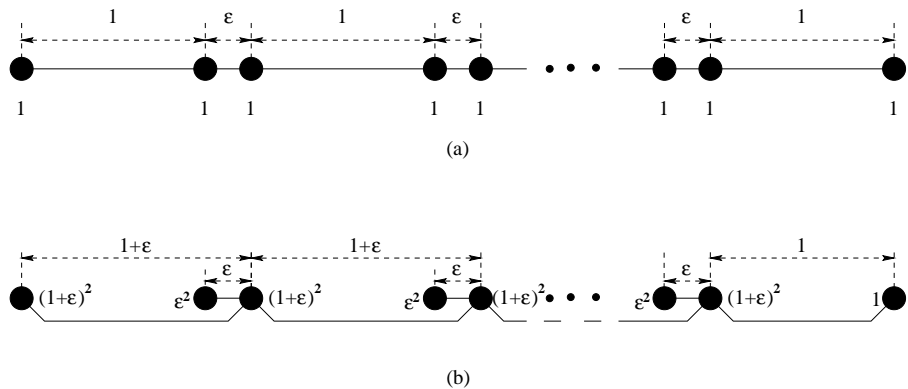,width=4.7in}} 
\caption{\label{f_2mst}
Tight example for the performance ratio of the MST algorithm.
In both cases, the solution is bidirected and the undirected
version of the arcs of the solution is given by solid edges.
For each vertex, its power in the solution is written next to it.
(a) The MST-based power assignment needs total power $2n$.
(b) Optimum power assignment has total power
$n(1+\epsilon)^2 + (n-1)\epsilon^2 + 1 \rightarrow n + 1$.
}
\end{figure}

The following known example (see Figure \ref{f_2mst})
 shows that the ratio of 2 for the MST algorithm is tight.
Consider $2n$ points located on a single line such that the distance
between consecutive points alternates between 1 and  $\epsilon<1$,
and let the cost function $c$ be the square of the Euclidean distance
Then the minimum spanning
tree MST connects consecutive neighbors and has power
$p(\MST) = 2n$.
On the other hand, the bidirected tree $T'$ with arcs connecting
each other node (see Figure \ref{f_2mst}(b)) has power equal
$p(T') = n(1+\epsilon)^2 + (n-1)\epsilon^2 + 1$. When
$n \rightarrow \infty$ and $\epsilon \rightarrow 0$, we obtain that
$p(\MST)/p(T') \rightarrow 2$.  On the other hand 
(argument taken from \cite{CH89}), for any input graph,
the power of the bidirected minimum spanning tree $T$ is at most
\[
p(T) =  \sum_{v \in V}  \max_{u | vu \in E(T)} c(vu) \leq
\sum_{v \in V}  \sum_{u | vu \in E(T)} c(vu) = 2 c(T) \leq 2 \opt
\]
where $\opt = p(\OPT)$ for an optimum solution $\OPT$ 
(the last inequality is from the introduction).

\begin{figure}
\centerline{\psfig{figure=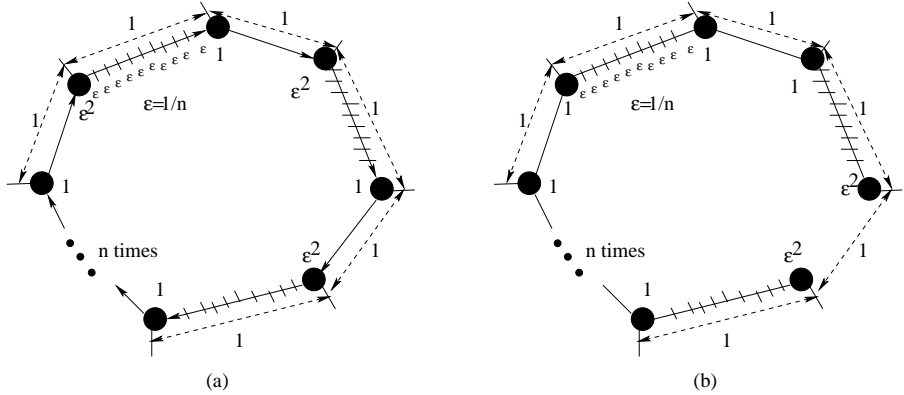,width=4.7in}}
\caption{\label{asymm}
Total power for Min-Power Strong Connectivity can be
half the total power for Min-Power Symmetric Connectivity.
In both cases, the solution is represented by solid segments
(arrows meaning not bidirected),
and next to each vertex its power is given.
(a) Minimum power assignment ensuring strong connectivity has total
power $n + n^2\eps^2=n+n^2\frac{1}{n^2}=n+1$.
 (b) Minimum power assignment ensuring symmetric connectivity has 
total power $(2n-2) + (n(n-1)+2)\eps^2  = 2n - 1-\frac{1}{n} + \frac{2}{n^2}$,
the solution is bidirected and the undirected
version of the arcs of the solution is given by solid segments.
}
\end{figure}

The example above may give intuition on how Power Assignment 
(and even more specifically, Min-Power Symmetric Connectivity
- the variant when $H$ must be bidirected mentioned in the introduction)
relates to the $k$-restricted Steiner trees, with stars
(trees of height 1) taking the place of restricted components. 
Another example  from \cite{ACMPTZ06}, (see Figure \ref{asymm}, and the
following paragraph), shows how Min-Power Strong Connectivity 
differs from Min-Power Symmetric Connectivity, and may give intuition how
Min-Power Strong Connectivity relates to Travelling Salesman, and also
Min-Cost Strong Connectivity and Min-Cost Two-Edge Connectivity
(a two-edge-connected graph has an edge orientation that makes it strongly
connected  - see for example Chapter 2,
written by A. Frank, of \cite{GrahamGL95}).
However we cannot think of direct reductions either way, and,
as we mention in Conclusions, the methods we use
only apply to certain instances of 
Min-Cost Strong Connectivity and Min-Cost Two-Edge Connectivity.

The power of a Min-Power Strong Connectivity optimum solution can be
almost half the power of a Min-Power Symmetric Connectivity optimum solution
for the same instance: we present a series of examples
illustrated in Figure \ref{asymm}.  The $n(n+1)$ vertices
are embedded in the plane in $n$ groups of $n+1$ points each.
Each group has two ``terminals"
(represented as thick circles in Figure \ref{asymm}),
and the $2n$ terminals are the corners of
a regular $2n$-gon with sides of length 1.
Each group has another $n-1$ equally spaced points
(dashes in Figure \ref{asymm}) on the line segment between the two terminals.
The cost function $c$ is the square of the Euclidean distance.
It is easy to see that a minimum power assignment
ensuring strong connectivity assigns a power of 1 to one thick
terminal in each group and a power of $\eps^2=(1/n)^2$ to all other points
 in the group - the arcs going clockwise.
The total power then equals $n+1$. For symmetric
connectivity it is necessary to assign power of 1 to all but
two of the thick points, and of $\eps^2$ to the remaining points,
which results in 
total power $(2n-2) + (n(n-1)+2)\eps^2  = 2n-1-1/n+2/n^2$.
Also, keep in mind that
the minimum spanning tree solution is a symmetric solution.

\section{The Approximation Algorithm}
\label{s_weighted}

This section is dedicated to proving the main result of this paper:

\begin{theorem}
There exists a polynomial time algorithm
for Min-Power Strong Connectivity with approximation ratio $1.85$.
\label{t_main}
\end{theorem}

The outline of the proof 
of Theorem \ref{t_main}, is as follows:
\begin{enumerate}
\item In the first subsection,
we present the greedy algorithm, preceded by necessary notation.

\item Then we establish (Lemma \ref{l_output_good}) that the algorithm's
output is strongly connected.

\item A second subsection gives the analysis.
It starts with the new lower bound, the fractional cover of tree
edges by stars.

\item Finally, Lemma \ref{l_zel} shows how the Greedy method of
\cite{RZ00} (applied with new parameters) combines the two lower bounds
(the one above, and the cost of the minimum spanning tree) 
to obtain the claimed approximation ratio.
\end{enumerate}

\subsection{The algorithm}

Our algorithm uses a greedy approach similar to \cite{Z96,RZ00,CFM07}.
Let $T$ be the undirected minimum spanning tree of $G$. 
The fact that $T$ has minimum cost will be not further used,
except to note $\opt \geq c(T)$,
where $\opt = p(\OPT)$ for an optimum solution $\OPT$.
Let $\tT$ be the bidirected version of $T$.

For $u \in V$ and $r \in \{c(uv) \; | \; uv \in E \}$,
let $S(u,r)$ be the directed star with center $u$ containing
all the arcs $uv$ with $c(uv) \leq r$; note that $r$ is the power of $S$,
also denoted by $p(S)$.
For a directed star $S$, let $E(S)$ be its set of arcs and
$V(S)$ be its set of vertices. 

For given $S(u,r)$, let $Q(u,r)$ be the set of edges $e$ of $T$
such that there exist $x,y \in V(S(u,r))$
with $e$ on the path from $x$ to $y$ in $T$.
Let $\hQ(u,r)$ be the set of arcs $e$ of $\tT$  
such that there exist $x \in V(S(u,r))$
with $e$ on the directed path from $u$ to $x$ in $\tT$;
it is easy to verify
that the undirected version of $\hQ(u,r)$ is $Q(u,r)$.

For a collection $\cA$ of directed stars $S(u_i,r_i)$, 
define $Q(\cA) = \bigcup_{ S(u_i,r_i) \in \cA } Q(u_i,r_i)$ and
$f(\cA) = \sum_{e \in Q(\cA)} c(e)$. We sometimes write $Q(S)$ instead of
$Q(\{S\}$.
The function $f(\cA)$ is known to be monotone and submodular  
(see an example in \cite{Schrijver03}, pages 768-769).
For $S = S(u,r)$, recall that $f_{\cA}(S) = f(\cA \cup \{S\}) - f(\cA) =
 \sum_{e \in Q(u,r) \setminus Q(\cA)} c(e) =  \sum_{e \in I_{\cA}(S)} c(e)$,
where $I_{\cA}(S)$ is defined to be those arcs of $\hQ(u,r)$ 
for which the undirected version is not in $Q(\cA)$.

The algorithm starts with $M=\tT$ as the set of arcs, and 
adds directed stars to collection $\cA$ (initially empty)
replacing arcs from $M$ to greedily reduce the sum of costs  
of the arcs in $M$ plus the sum of the powers of the stars in $\cA$.
For intuition, we mention
that this sum is our upper bound on the power of the algorithm's output.
To simplify later proofs, the algorithm makes changes 
(adding directed stars and removing arcs from $M$)
even if our sum stays the same. Assume below that $0/0=1$. To be precise:

\begin{tabbing}
\ \ \ \= \ \ \= \ \ \= \  \ \= \ \ \  \\
Algorithm {\bf Greedy:} \\
\> $\cA \leftarrow \emptyset$, $M \leftarrow \tT$ \\
\> While ($f(\cA) < c(T)$ ) do\\
\> \> $(u,r) \leftarrow \mbox{argmax}_{(u',r')} f_{\cA}(S(u',r'))/r' $ \\
\> \> $ M \leftarrow M \setminus I_{\cA}(S(u,r)) $ \\
\> \> $ \cA \leftarrow \cA \cup \{S(u,r)\}$ \\
\> Output $ \bigcup_{S \in \cA} E(S) \cup M$
\end{tabbing}

\bigskip

\begin{figure}[ht]
\psfrag{x}{\tiny{$x$}}
\psfrag{y}{\tiny{$y$}}
\psfrag{u}{\tiny{$u$}}
\psfrag{v}{\tiny{$v$}}
\begin{minipage}[b]{0.3\linewidth}
\centering
\includegraphics[scale=0.4]{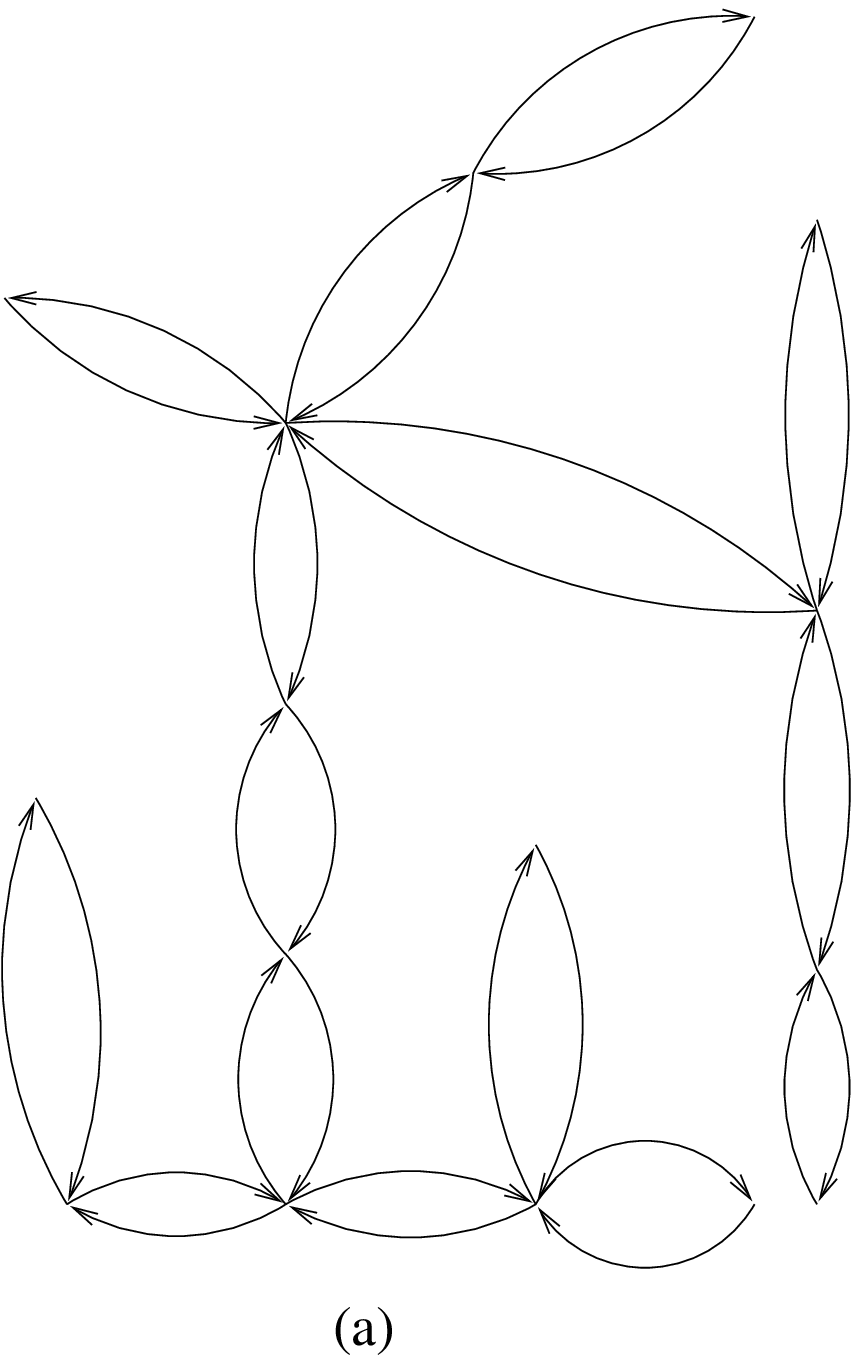}
\end{minipage}
\begin{minipage}[b]{0.3\linewidth}
\centering
\includegraphics[scale=0.4]{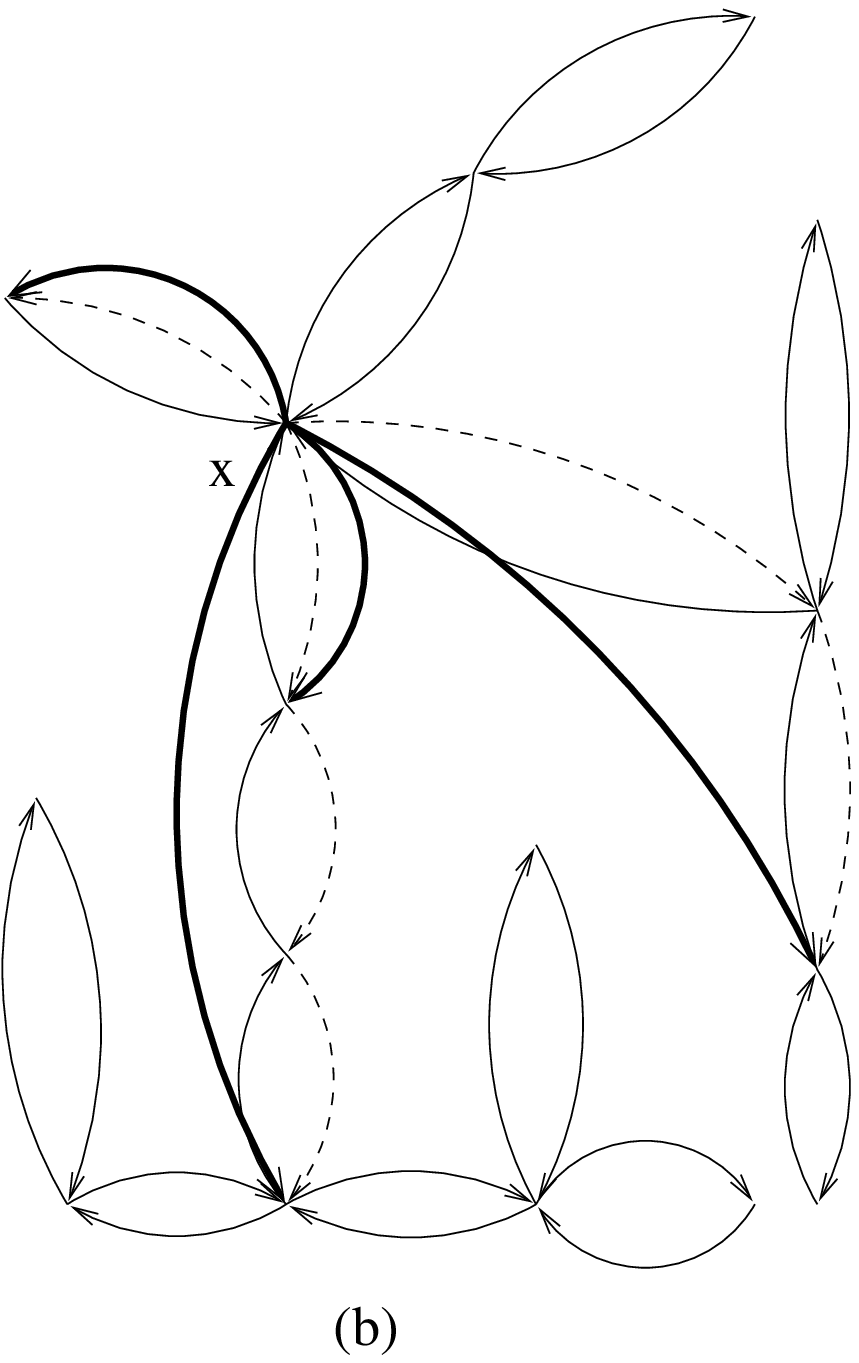}
\end{minipage}
\begin{minipage}[b]{0.3\linewidth}
\centering
\includegraphics[scale=0.4]{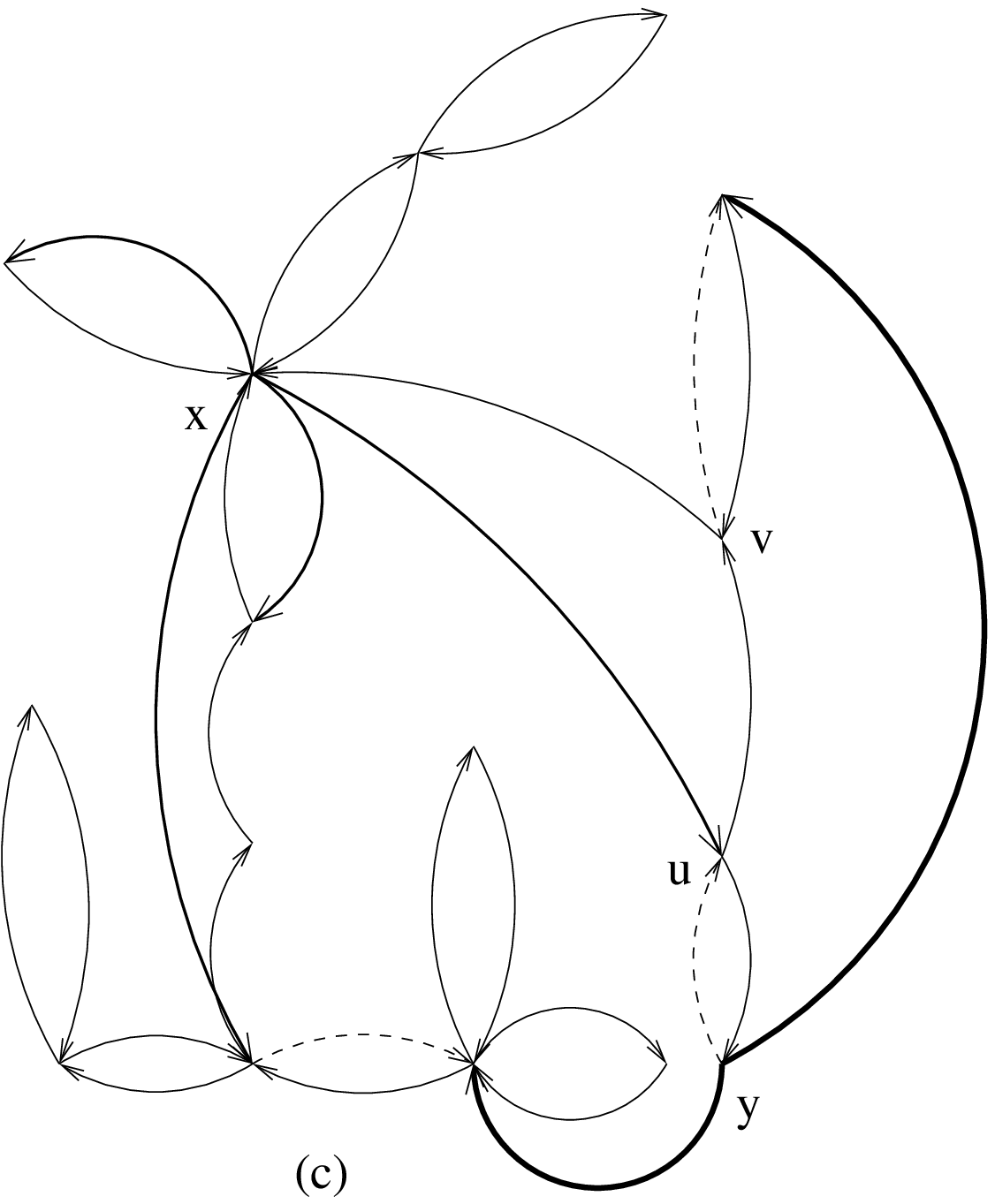}
\end{minipage}
\caption{
Costs are not relevant here.  (a) Initial $M$.  (b) Solid arcs give 
$\bigcup_{S \in \cA} E(S) \cup M$ after adding to $\cA$
 the star $S_1$ (with thick solid arcs) centered at $x$.
The arcs removed from $M$ are dashed.
(c) Solid arcs give 
$\bigcup_{S \in \cA} E(S) \cup M$ after adding 
to $\cA$ the star $S_2$ (with thick solid arcs) centered at $y$.
The arcs removed from $M$ are dashed.
Note that the algorithm does not remove arc $uv$.
Once an arc from a pair of antiparallel arcs of $\tT$ 
is removed, the algorithm keeps the other arc, since
there are cases (as in Figure \ref{f_again})
 when not all arcs of $\hQ(u,r)$ can be removed
while keeping strong connectivity, and benefiting from
removing arcs whose antiparallel arc has already been removed from $\tT$
(when possible)
destroys the ``submodularity" implictly 
needed in the approximation ratio proof.
}
\label{f_algo}
\end{figure}

\begin{figure}[ht]
\psfrag{x}{\tiny{$x$}}
\psfrag{y}{\tiny{$y$}}
\psfrag{u}{\tiny{$u$}}
\psfrag{v}{\tiny{$v$}}
\begin{minipage}[b]{0.3\linewidth}
\centering
\includegraphics[scale=0.4]{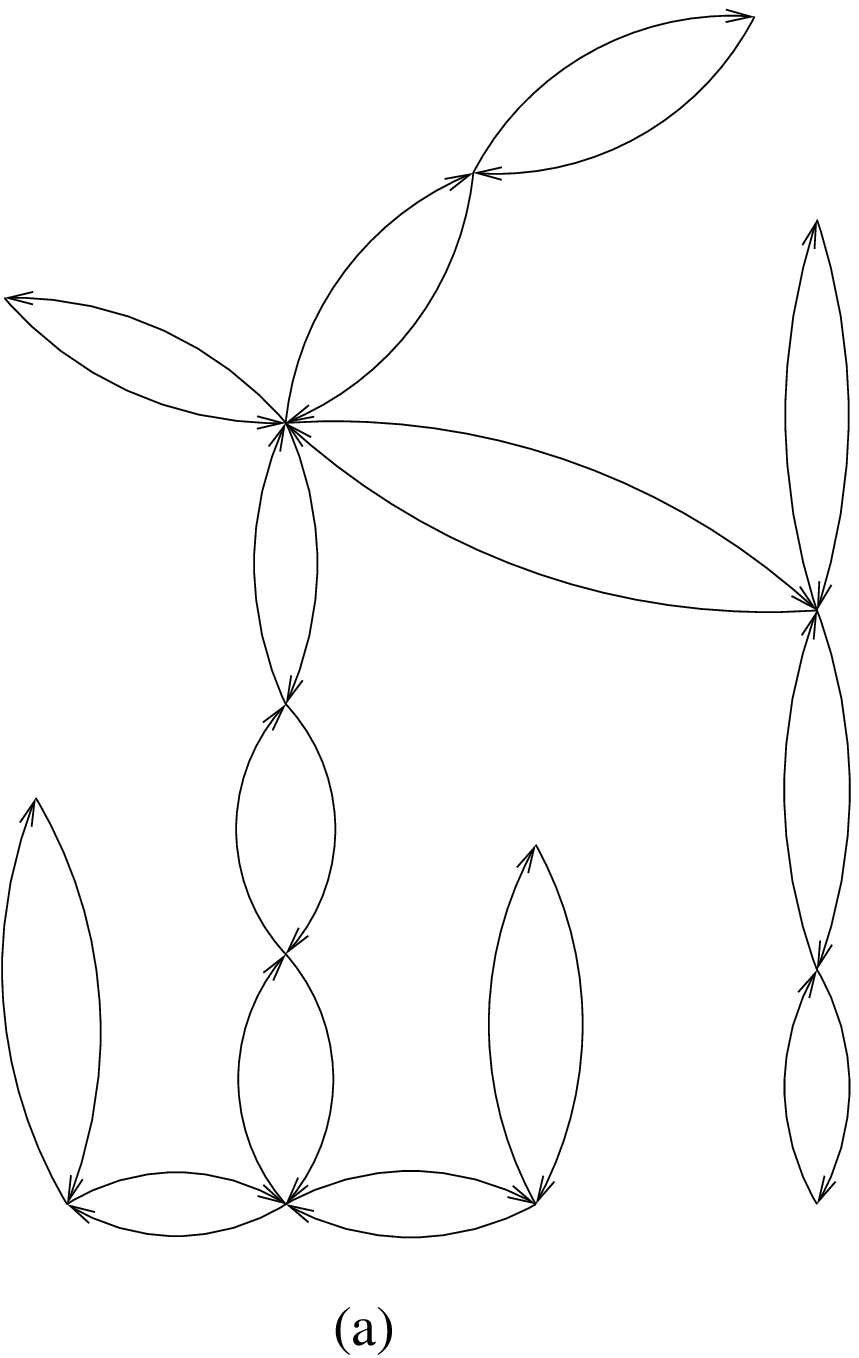}
\end{minipage}
\begin{minipage}[b]{0.3\linewidth}
\centering
\includegraphics[scale=0.4]{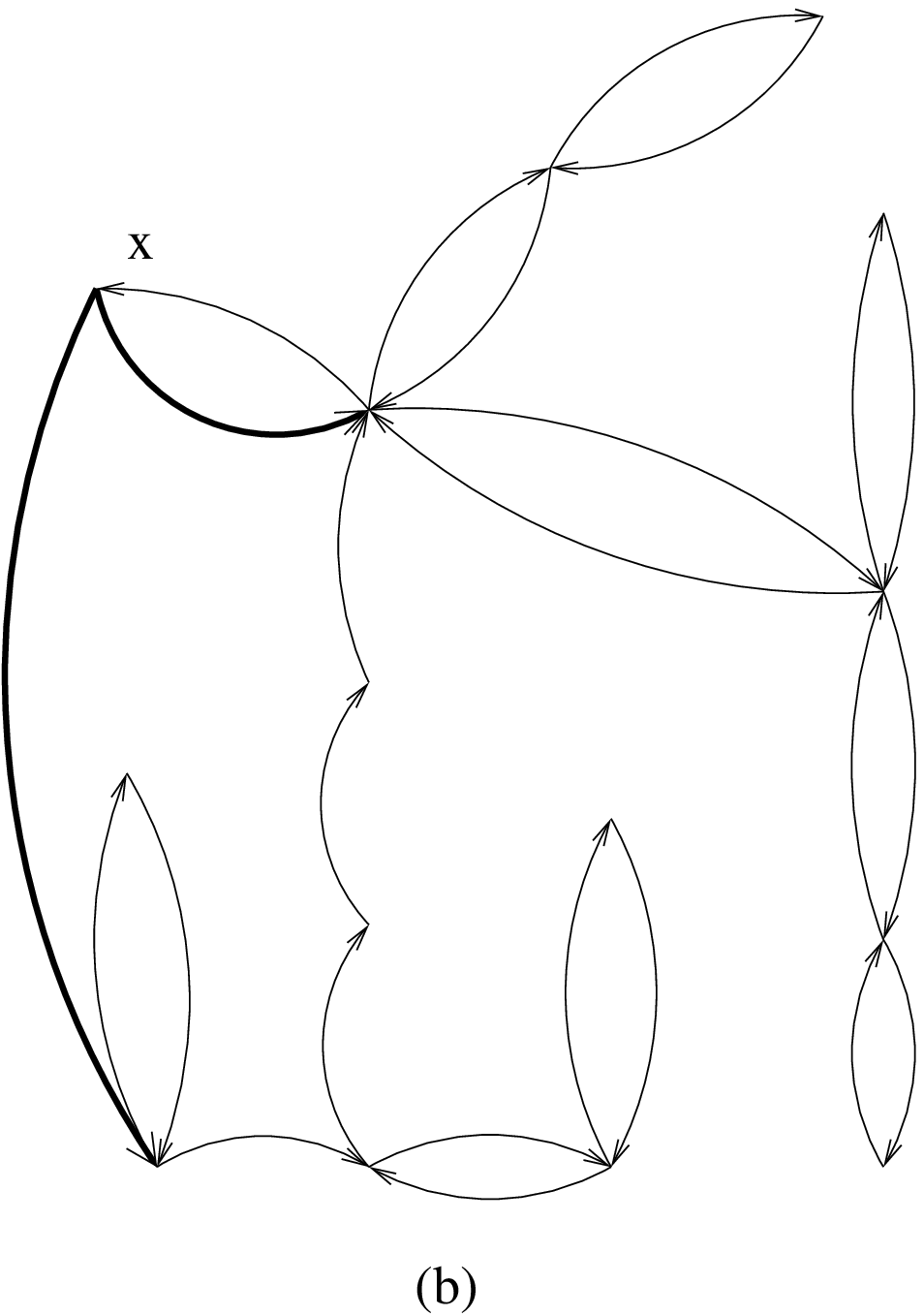}
\end{minipage}
\begin{minipage}[b]{0.3\linewidth}
\centering
\includegraphics[scale=0.4]{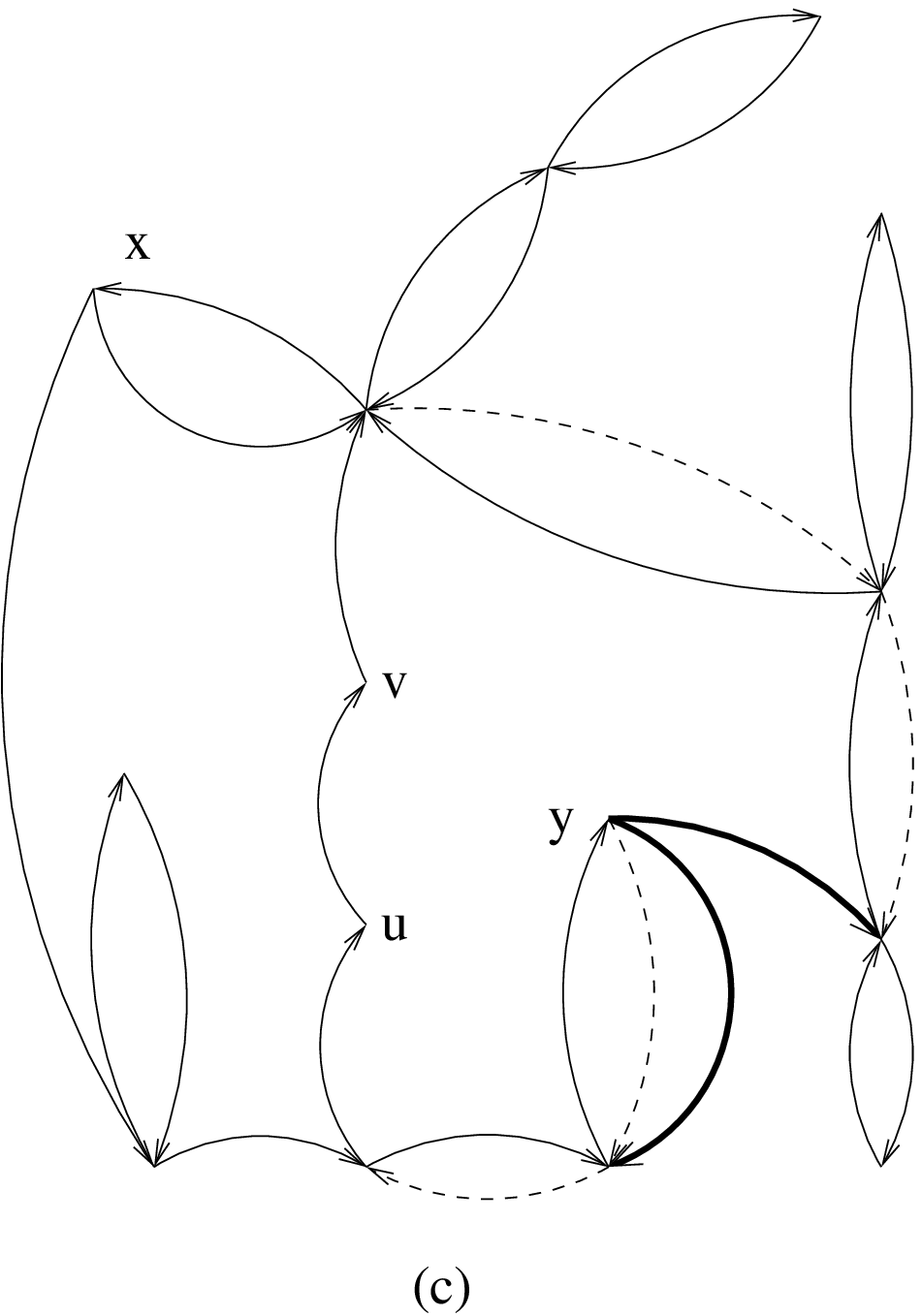}
\end{minipage}
\caption{
Costs are not relevant here.  (a) Initial $M$.  
(b) $\bigcup_{S \in \cA} E(S) \cup M$ after adding to $\cA$
 the star $S_1$ (with thick solid arcs) centered at $x$,
and removing arcs from $M$.
(c) Solid arcs give 
$\bigcup_{S \in \cA} E(S) \cup M$ after adding 
to $\cA$ the star $S_2$ (with thick solid arcs) centered at $y$.
The arcs removed from $M$ are dashed.
Note that the algorithm cannot remove arc $uv$ even though
$uv \in \hQ(S_2)$, since $u$ and $v$ become disconnected.
}
\label{f_again}
\end{figure}

Each of figures \ref{f_algo} and \ref{f_again} shows
 two iterations of the algorithm.
For intuition, we mention that this algorithm ``covers" undirected 
edges of the minimum spanning tree
by ``stars" and when implemented, it is a variant of
Chvatal's \cite{Chvatal79} greedy algorithm for Set Cover.

\begin{fact}
\label{obs}
Note that, unless $f(\cA) = c(T)$, a star $S(u,r)$ always exists for which 
$f_{\cA}(S(u,r)) > 0$ and $f_{\cA}(S(u,r))/r \geq 1$.
Indeed, as long as a pair of antiparallel arcs $e'$ and $e''$ are
in $M$, 
we can pick as next star $S(u,r)$ the one
given by $u$ being the tail of $e'$ and $r = c(e')$.
\end{fact}

Thus, as written, the algorithm can have iterations that do not change
the output, i.e. above the star $S(u,r)$ above could
have just the edge $e'$ and
be added to $\cA$ while $e'$ is removed from $M$. 

\begin{lemma}
The output of {\bf Greedy} is a spanning strongly connected subgraph of $G$.
\label{l_output_good}
\end{lemma}
\proof
We prove the following invariant:
$X := \bigcup_{S \in \cA} E(S) \cup M$
gives a spanning strongly connected subgraph whenever the {\bf while}
condition is checked by the algorithm. Moreover,
suppose we remove from $T$ all edges for which both
antiparallel arcs appear in $M$, splitting $T$ in components
with vertex sets $T_i$, for some range of $i$.
We prove that
for every $i$ and every $x,y \in T_i$, there exists a
directed  path $P$ from $x$ to $y$  using only vertices of $T_i$
and arcs from $X$.

\begin{figure}[bht]
\begin{center}\leavevmode%
\psfrag{x}{$x$}
\psfrag{y0}{$u_0$}
\psfrag{y1}{$y_1$}
\psfrag{y2}{$y_2$}
\psfrag{z0}{$z_0$}
\psfrag{z1}{$z_1$}
\psfrag{z2}{$z_2$}
\psfrag{z3}{$z_3$}
\psfrag{z4}{$z_4$}
\psfrag{u}{$u$}
\psfrag{u1}{$u_1$}
\scalebox{.7}{
\includegraphics{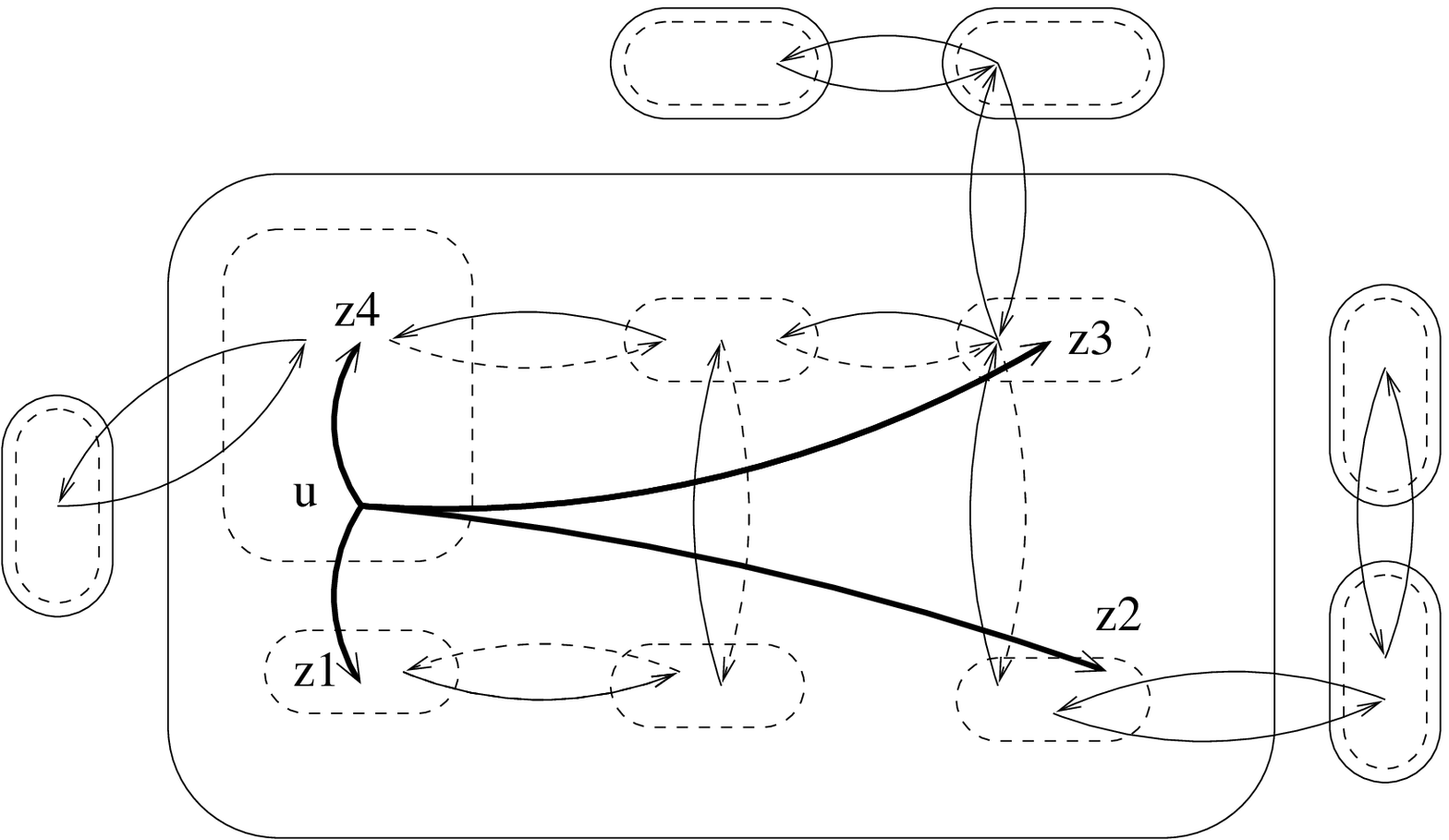}
}
\end{center}
\caption{
Rounded rectangles show the components $T_i$, dashed before $S$
is added to $\cA$ (the split of $T$ by $M$) and solid afterward 
(the split of $T$ by $M'$).
Arcs of $M$ crossing from one component to another are given by thin arcs,
$S$ by the four thick arcs $uz_1,uz_2,uz_3,uz_4$,
and the dashed arcs are those removed from $M$ when $S$ is added.
}
\label{f_invariant}
\end{figure}

Proving the invariant holds is done as always 
by induction on the number of iterations.
The invariant is true before the first iteration, when each $T_i$
has just one vertex, so consider the moment when
a star $S = S(u,r)$ is added to $\cA$. Figure \ref{f_invariant} 
may provide intuition.
We add arcs $uz$, for $z \in V(S) \setminus \{u\}$,
while removing from $M$ (and from  $X$)
  arcs $xy$ if $yx \in M$ and there is some $z$ such that
$xy$ is on the directed simple path from $u$ to $z$ in $\tT$.
The same effect is obtained if we do this change for 
each $z \in V(S) \setminus \{u\}$ one after the other,
instead of all such $z$ at the same time.

Let $P$ be the simple path in $T$ from $u$ to $z$, and let
$x_i y_i$, for $1 \leq i \leq k$,  
be, in order, the arcs of $M$ on $P$ such that also $y_i x_i \in M$.
Thus the change to $X$ consists of adding the arc $uz$ and
removing all arcs $x_i y_i$; note that if $k=0$ no arc is removed
and our induction step is complete. 
Let $M' = M \setminus \{x_1 y_1, \ldots, x_k y_k\}$ and
$X' = X \setminus 
\{x_1 y_1, \ldots, x_k y_k\} \cup \{uz\}$.
We need to show that $X'$ and $M'$ satisfy the conditions
from the induction hypothesis.

Let us split $T$ into components by removing all
the undirected edges $xy$ with both antiparallel arcs  $xy$ and $yx$ in $M$; in
particular all $x_i y_i$, for $1 \leq i \leq k$,
resulting in components $T_i$.

By induction,  $X$ contains the following directed paths:
$P_1$ from $x_1$ to $u$, $P_2$ from $x_2$ to $y_1$, \ldots, 
$P_k$ from $x_k$ to $y_{k-1}$, and $P_{k+1}$ from $z$ to $y_k$,
and each of these paths stays in the same component $T_i$ of 
the split of $T$ by $M$ - as in the previous paragraph.
Thus none of these paths uses $x_i y_i$ or $y_i x_i$.
Putting together these paths, the arcs $y_i x_i$, for $1 \leq i \leq k$,
and the arc $uz$, we have a directed cycle $C$
containing none of the arcs  $x_i y_i$, for $1 \leq i \leq k$.
Any arc removed can be replaced, when discussing connectivity,
with a path around the cycle $C$, and so $(V,X')$ is strongly connected,
as required.

We now split $T$ into components by removing all
the undirected edges $xy$ with both $xy$ and $yx$ in $M'$,
obtaining components $T'_i$. 
Note that none of $P_i$, $1 \leq i \leq k+1$, from above, has an arc
with endpoints in two distinct components $T'_i$ (as $T'_i$ is the union
of several $T_j$). 
As all the edges on the path from $u$ to $z$ in $T$
do not have anymore both antiparallel arcs in $M'$,
all the vertices on this path including $u,z$ and all $x_i,y_i$
are in the same component $T'_j$ of the split of $T$ by $M'$.
Thus all the arcs of $C$ have their endpoints in the
same component of the split of $T$ by $M'$.

We prove that
for every $i$ and every $x,y \in T'_i$, there exists a
directed  path $P'$ from $x$ to $y$  using only vertices of $T'_i$
and arcs from $X'$. 
First, let us describe a path $P$ from $x$ to $y$ using only arcs of $X$:
find the path from $x$ to $y$ in $T$,  and let 
$z_i w_i$, for $1 \leq i \leq q$,  
be, in order, the arcs of $M$ on $P$ such that also $z_i w_i \in M$.
When $q=0$, $x,y \in T_j$ for some $j$
(same component of the split of $T$ by $M$) and by
induction, a path $P$ from $x$ to $y$ exists in $X$ using only vertices
inside $T_j$. We pick $P'=P$, and indeed $P'$ only uses arcs of $X'$
since the arcs of $X \setminus X'$ (same set as $M \setminus M'$)
cross from one component to another of the split of $T$ by $M$.
Assume now $q>0$. Notice that all unordered pairs $z_i w_i$
belong in the set of unordered pairs $x_j y_j$ on the simple path from
$u$ to $v$ mentioned earlier, or else we cannot have that $x$ and $y$
belong to the same $T'_i$ of the split of $T$ by $M'$.
Also, by induction,  $X$ contains the following directed paths:
$P_1$ from $x$ to $z_1$, $P_2$ from $w_1$ to $z_2$, \ldots, 
$P_q$ from $w_{q-1}$ to $z_q$, and $P_{q+1}$ from $w_q$ to $y$,
and each of these paths stays in the same component of the split of $T$ by $M$.
Thus none of these paths uses $z_i w_i$ or $w_i z_i$.
Next, obtain $P'$ by replacing in $P$,
if necessary, arcs of $X \setminus X'$ (same set as $M \setminus M'$)
by arcs of $C$, staying, as shown in the previous paragraph,
in the same component of $T$ split by $M'$.
Note that $P'$ indeed uses only vertices of $T'_i$.
This completes the induction step.  \qed

\subsection{Approximation ratio analysis}

For a collection $\cA$ of directed stars $S(u_i,r_i)$, 
define $w(\cA) = \sum_{ S(u_i,r_i) \in \cA } r_i$, 
the total power used by the stars in $\cA$.

\begin{lemma}
Let $\cB$ be a an arbitrary collection of stars,
and $T$ be an arbitrary spanning tree.
There exist non-negative coefficients $x_S$ (over the collection of
all possible stars $S(u,r)$) such that
$\sum_{S} x_S f_{\cB} (S) \geq c(T) - f(\cB)$ 
and $\sum_{S} x_S p(S)  \leq (1/2) \opt$.
\label{l_fractional}
\end{lemma}


\proof
We assume that $c(T) - f(\cB) > 0$, or else $x_S = 0$ for all $S$ will do.
Assign $x_S = (1/2)$ for  every star of $\OPT$, and $x_S = 0$ otherwise.
Therefore $\sum_{S} x_S p(S)   =  (1/2) \opt$.

Recall that 
$Q(\cB) = \bigcup_{ S(u_i,r_i) \in \cB } Q(u_i,r_i)$ and
let $e \in T \setminus Q(\cB)$. 
If we remove $e$ from $T$, we create two subtrees
$T_u$ and $T_v$, where $u$ and $v$ are the endpoints of $e$.
$\OPT$, being strongly connected,
has at least one star $S_u$ with the center in $V(T_u)$ and one of 
its other vertices in $V(T_v)$. Then $e \in Q( \cB \cup \{S_u\} )$.
Similarly, $\OPT$  
has at least one star $S_v$ with the center in $V(T_v)$ and one of 
its other vertices in $V(T_u)$. Then $e \in Q( \cB \cup \{S_v\} )$.
Note that $S_v \neq S_u$ (centers in disjoint vertex sets).

We have:
\begin{eqnarray*}
\sum_{S} x_S f_{\cB} (S)  & = &  
\sum_S x_S \sum_{e \in Q(\cB \cup \{S\}) \setminus Q(\cB)} c(e) \\
& = & \sum_{e \in T} c(e) 
\sum_{S \; | \; e \in Q(\cB \cup \{S\}) \setminus Q(\cB)} x_S  \\
& = & \sum_{e \in T \setminus Q(\cB)} c_e \sum_{ S \; | \; e \in Q(S)} x_S  \\
& \geq & \sum_{e \in T \setminus Q(\cB)}  c_e  = c(T) - f(\cB),
\end{eqnarray*}
where the inequation is given by the two distinct stars $S_u$ and $S_v$
described above for every edge $e \in T \setminus Q(\cB)$.
This completes the proof.
\qed

The example from Figure \ref{f_2mst} shows that a constant
better than $1/2$ is not possible above:
with $\cB$ being empty, we have $c(T)$ circa $n$, and for each star $S$
of optimum, $p(S)$ is circa $1$ and $f(S)$ circa $2$.

Now we need the following lemma, whose proof is obtained from 
Robins-Zelikovsky as presented in \cite{GroplHNP01} by changing 
what quantities represent and some parameters,
together with a ``fractional cover" idea from 
the journal submission version of \cite{BGRS10}.

\begin{lemma}
Assuming that for the minimum spanning tree $T$,
and for any collection of stars $\cB$,
there exist non-negative coefficients $\left( x_S \right)$ such that
$\sum_{S} x_S f_{\cB} (S) \geq c(T) - f(\cB)$ 
and $\sum_{S} x_S p(S)  \leq \alpha \opt$,
where $\opt$ is the power of the optimum solution and $\alpha<1$, 
the algorithms' output has power at most  $\beta \opt$
where $\beta = 1 + \alpha + \alpha \ln (1/\alpha)$. 
\label{l_zel}
\end{lemma}
\proof
First, if $c(T) \leq \alpha \opt$, then before any improvement 
we have a solution of cost at most $2 \alpha \opt$ and $2 \alpha < \beta$.
Thus in the following we assume $\opt \geq c(T) > \alpha \opt$
(the first inequality is due to $T$ being a minimum spanning tree).

Note that at the end of the algorithm, 
$M$ contains exactly one of the two antiparallel
arcs for each edge of $T$. Then,
for the final collection of stars $\cA$, the output $H$ satisfies
\begin{equation}
\label{e_opt}
p(H) \leq c(T) + w(\cA)
\end{equation}
as it follows by summation over $u \in V$ from
\[
p_H(u) = max_{uv \in H} c(uv) \leq \sum_{uv \in M} c(uv) + 
\sum_{S \in \cA} p(S),
\]
which holds for every vertex $u$ 
(recall that $E(H)  = \bigcup_{S \in \cA} E(S) \cup M$).
 

Let $S_1, S_2, \ldots, S_q$ be the stars picked by our algorithm
and  let $\cA_i$, for $1 \leq i \leq q$ be the collection of the first
$i$ stars; also let for convenience $\cA_0$ be the empty collection.
For $1 \leq i \leq q$, let $p_i = p(S_i)$, 
and let $f_i = f_{\cA_{i-1}}(S_i)$.
Note then that for all $i$, since 
$f_{\cA_{i-1}}(S_i) =  f(\cA_{i}) - f(\cA_{i-1})$,
we have $f(\cA_{i}) = \sum_{j=1}^{i} f_j$.

If $f_i = 0$, then $p_i = 0$ and Equation \ref{e_p_i} below holds. Otherwise,
the greedy choice of the algorithm and the assumptions of the theorem
for $\cB = \cA_{i-1}$ give:

\begin{equation}
p_i \leq f_i \frac{\alpha \opt}{c(T) - \sum_{j=1}^{i-1} f_j}.
\label{e_p_i}
\end{equation}
Define the function $g: [0 .. c(T)] \rightarrow [0 .. 1]$ by
$g(x) = \alpha \opt/(c(T) - x)$ for $x \leq c(T) - \alpha \opt$,
and $g(x) = 1$ for $x > c(T) - \alpha \opt$.
Then from Equation \ref{e_p_i} and Fact \ref{obs}
(that $p_i \leq  f_i$), we obtain:
\[
p_i \leq \int_{\sum_{j=1}^{i-1} f_j}^{\sum_{j=1}^{i} f_j} g(x) dx.
\]

\begin{figure}[bht]
\begin{center}\leavevmode%
\psfrag{f1}{$f_1$}
\psfrag{f2}{$f_2$}
\psfrag{fq}{$f_q$}
\psfrag{fi}{$f_i$}
\psfrag{fi+}{$f_{i+1}$}
\psfrag{fi++}{$f_{i+2}$}
\psfrag{cT}{$c(T)$}
\psfrag{cT-a}{$c(T) - \alpha \opt$}
\scalebox{.4}{
\includegraphics{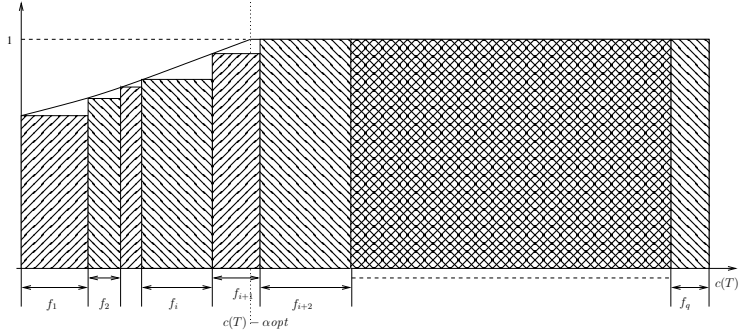}
}
\end{center}
\caption{
The function $g(x)$ is given by the solid curve.
$\sum_{i=1}^q p_i$ is the shaded area, in rectangles of width $f_i$
and height 
$\frac{\alpha \opt}{c(T) - \sum_{j=1}^{i-1} f_j}$.
In this particular picture, $\alpha \opt = (2/3) c(T)$ and therefore
the integral is circa $0.94 c(T)$.
}
\label{f_integral}
\end{figure}

Therefore (see Figure \ref{f_integral}):

\begin{eqnarray*}
\sum_{i=1}^q p_i  & \leq & \int_0^{c(T)} g(x) dx = 
\int_0^{c(T) - \alpha \opt } \frac{\alpha \opt}{c(T) - x} dx  + 
\int_{c(T) - \alpha \opt }^{c(T)}  1 dx   \\
& = & ( - \alpha \opt) \ln (c(T)-x) \Big \arrowvert_{0}^{c(T) - \alpha \opt} + 
	\left(c(T) - ( c(T) - \alpha \opt)  \right) \\
& = &  ( - \alpha \opt) \left( \ln (c(T) - (c(T) - \alpha \opt) ) - \ln c(T) \right)  + \alpha \opt
 = \alpha \opt \left( 1 +  \ln \frac{c(T)}{\alpha \opt} \right)
\end{eqnarray*}

Using this and $c(T) \leq \opt$ and
Equation \ref{e_opt} (recall that $w(\cA) = \sum_{i=1}^q p_i$), we obtain
that the power of the output is at most
\[
c(T) + \alpha \opt \left( 1 +  \ln \frac{c(T)}{\alpha \opt} \right)
\leq \opt \left( 1 +  \alpha +  \alpha \ln (1/\alpha) \right) 
\]
finishing the proof.  
\qed

Based on Lemmas \ref{l_output_good}  and
\ref{l_fractional}, Theorem \ref{t_main} follows immediately from
the fact that $\alpha = 1/2$ makes $\beta \leq 1.85$.
Note also that $\alpha < 1$ implies $\beta < 2$, which
follows from $\alpha ( 1 + \ln (1/\alpha) ) < 1$, 
which is equivalent to $ \ln (1/\alpha) < 1/\alpha - 1$, a fact that holds
for all $\alpha < 1$. 
\cite{C10} had $\alpha = 7/8$.

\section{Linear Programming Relaxation}

While not improving the approximation ratio of Greedy,
this may allow for further LP-based algorithm.
The following natural Integer Linear Program is called {\em IP2} as in
\cite{CQ12}. We adapted the notation,
and have variables $y_S$ for every arc $S = S(u,r)$.
The idea is that $y_S$ being 1 ``represents"  that $S$ is 
star of the optimum solution.
We say that star $S = S(u,r)  \in \delta^-(X)$,
for $ X \subset V, \emptyset \neq X \neq V$
iff $u \not \in X$ and $V(S) \cap X \neq \emptyset$.

\begin{equation*}
\mbox{ minimize } \sum_{S} y_S p(S)
\mbox{  subject to}
\end{equation*}
\begin{eqnarray}
\label{c_ip2_cross}
\sum_{S \in \delta^-(X)} y_S \geq 1 \; & \forall \; X \subset V, 
        \emptyset \neq X \neq V \\
\label{c_ip2_x_nonn}
y_S \geq 0  \; & \forall \; S\\
\label{c_ip2_int}
y_S \in \ZZ \; & \forall \; S
\end{eqnarray}

{\bf LP2} is the linear relaxation of IP2; that is,
the linear program given by exactly the same constraints except
the last one. 
LP2 has exponentially many ``cut"
constraints, but when we replace them by ``flows", 
 we have $O(mn^2)$ non-zero entries in the matrix, as shown in \cite{CQ12}.
Thus it can be solved in polynomial-time.

Let $\opt^*$ be the optimum of the linear program LP2 for a given instance.
Then clearly $\opt^* \leq \opt$ and \cite{CQ12} proves that 
$\opt \leq 2 \opt^*$.
We do better here. First, Lemma \ref{l_fractional} has a straightforward
adaptation:

\begin{lemma}
Let $\cB$ be a an arbitrary collection of stars,
and $T$ be an arbitrary spanning tree.
There exist non-negative coefficients $x_S$ (over the collection of
all possible stars $S(u,r)$) such that
$\sum_{S} x_S f_{\cB} (S) \geq c(T) - f(\cB)$ 
and $\sum_{S} x_S p(S)   \leq (1/2) \opt^*$.
\label{l_fractional_lp}
\end{lemma}

\proof
We assume that $c(T) - f(\cB) > 0$, or else $x_S = 0$ for all $S$ will do.
Let $y_S$ (for all $S = S(u,r)$) be a an optimum solution of LP2.
Assign $x_S = (1/2) y_S$ for all $S$; therefore
$\sum_{S} x_S p(S)   =  (1/2) \opt^*$.

Recall that 
$Q(\cB) = \bigcup_{ S(u_i,r_i) \in \cB } Q(u_i,r_i)$ and
let $e \in T \setminus Q(\cB)$. 
If we remove $e$ from $T$, we create two subtrees
$T_u$ and $T_v$, where $u$ and $v$ are the endpoints of $e$.
Constraint \ref{c_ip2_cross} gives
\[
\sum_{S \in \delta^-( V(T_u) )} x_S \geq 1/2
\]
and
\[
\sum_{S \in \delta^-( V(T_v) )} x_S \geq 1/2
\]
Note that $\delta^-( V(T_u) )$ and  $\delta^-( V(T_u) )$
are disjoint sets since a star in the first set has
its center in $V(T_v)$, while
a star in the second set has its center in $V(T_u)$.
Moreover, any star $S \in \delta^-( V(T_u) ) \cup \delta^-( V(T_v) )$
has $e \in Q(S)$. Thus
\begin{equation}
\label{e_frac_cover}
\sum_{S \; | \; e \in Q(S) } x_S \geq 1
\end{equation}
 
From now on we coppied from the previous proof:
\begin{eqnarray*}
\sum_{S} x_S f_{\cB} (S)  & = &  
\sum_S x_S \sum_{e \in Q(\cB \cup \{S\}) \setminus Q(\cB)} c(e) \\
& = & \sum_{e \in T} c(e) 
\sum_{S \; | \; e \in Q(\cB \cup \{S\}) \setminus Q(\cB)} x_S  \\
& = & \sum_{e \in T \setminus Q(\cB)} c_e \sum_{ S \; | \; e \in Q(S)} x_S  \\
& \geq & \sum_{e \in T \setminus Q(\cB)}  c_e  = c(T) - f(\cB),
\end{eqnarray*}
where the inequation is from (\ref{e_frac_cover}).
This completes the proof.
\qed

The example from Figure \ref{f_2mst} again shows that a constant
smaller than $1/2$ is not possible in Lemma \ref{l_fractional_lp}:
with $\cB$ being empty, we that for each star $S$
with $f(S) > \eps$, $p(S) \geq  (1/2 - \eps) f(S)$.
Thus, if $\sum_{S} x_S f (S) \geq c(T)$, then  
$\sum_{S} x_S p(S) \geq (1/2 - \eps) c(T) 
\geq (1/2 - \eps) (1 - \gamma) \opt \geq 
(1/2 - \eps - \gamma) \opt^*$,
where we used that the example can be made to have 
$\opt (1 - \gamma) \leq c(T)$, for any $\gamma > 0$.
As $\gamma$ and $\eps$ can be made arbitraly small,
we can see that indeed a constant smaller than $1/2$ is not possible.

It is also proven in Section II of \cite{CQ12} that $c(T) \leq \opt^*$;
then Lemma \ref{l_zel} goes through with $\opt^*$ instead of $\opt$.
Therefore we conclude that the output of Greedy is at most $1.85 \opt^*$.

\section{Conclusions}

This work greatly simplifies and at the same time improves our 
earlier work \cite{C10}.
Instead of Greedy we could have used the 
the iterative randomized rounding of
 Byrka et. al \cite{BGRS10} for Steiner Tree,
with the same approximation ratio.
However, we do not see further improvements coming from
using their full range of techniques,
 since we do not see the equivalent of the concept of ``loss"
used explicitly by \cite{RZ00} and implicitly by  \cite{BGRS10}.

\section{Acknowledgments}

We would like to thank Alexander Zelikovsky for a streamlined explanation
of his algorithms,
and to Gianpiero Monaco and Anna Zych for a group reading of \cite{BGRS10}
at the University of Warsaw.

\end{document}